\documentclass[9pt,twocolumn,twoside]{opticajnl_arxiv}

\journal{opticajournal} 

\setboolean{shortarticle}{true}

\usepackage{lineno}

\usepackage{siunitx}

\usepackage{todonotes}
\usepackage{cancel}
\usepackage{xcolor}

\usepackage{printlen} 
\uselengthunit{in}    

\title{Programmable Focal Elongation and Shaping of High-Intensity Laser Pulses using Adaptive Optics}

\author[1]{Peter Blum}
\author[1]{Anna Puchert}
\author[1]{Emily Archer}
\author[1]{S\"{o}ren Jalas}
\author[2]{Spencer W. Jolly}
\author[1,$\dagger$]{Jens Osterhoff}
\author[1]{Wim P. Leemans}
\author[1]{Manuel Kirchen}
\author[1]{Andreas R. Maier}
\author[1, *]{Rob J. Shalloo}

\affil[1]{Deutsches Elektronen-Synchrotron DESY, Notkestraße 85, 22607 Hamburg, Germany}
\affil[2]{Service OPERA-Photonique, Université libre de Bruxelles (ULB), Brussels, Belgium}

\affil[$\dagger$]{Now at Lawrence Berkeley National Laboratory}
\affil[*]{rob.shalloo@desy.de}

\begin{abstract} 
Controlling the intensity distribution of laser pulses in the focal region is essential for optimizing optically generated plasma waveguides and enabling advanced plasma acceleration techniques, including dephasingless wakefield acceleration.
Here, we present a method for programmatic structuring of the high-intensity focal region of a standard off-axis parabolic mirror, extending the length of this region well beyond the Rayleigh length and enabling control over the longitudinal intensity distribution.
The theoretical framework is validated through numerical simulations and experimental measurements.
Further, we demonstrate the use of this technique in an existing plasma accelerator system using readily available hardware components.
Finally, we illustrate the potential application of this method to multi-GeV laser plasma acceleration and the generation of flying foci, research areas which would significantly benefit from improved programmatic structuring of high-intensity laser pulses. 
\end{abstract}

\setboolean{displaycopyright}{false}

\begin{document}
\maketitle

Extended depth of focus optics, or \emph{axioptics}, are becoming increasingly important in many areas of high-power laser-matter interactions. 
Rather than focusing rays of light to a single point, like an off-axis parabolic mirror (OAP), these optics focus to a line segment along the optical axis, allowing for the generation of extended regions of high laser intensity. 
Optics for generating such intensity structures include the axicon \cite{McLeod1954}, the axilens \cite{Davidson1991,Sochacki1992} and the more recently proposed axiparabola \cite{Smartsev2019,Oubrerie2022}.

Axioptics are routinely used for the formation of optical waveguides to prevent diffraction of the drive laser pulse in a laser-plasma accelerator (LPA), which is a fundamental requirement for efficient acceleration to multi-GeV energies \cite{Oubrerie2022,Miao2022,Picksley2024}. 
They have also been proposed as a key optical element for applying flying focus techniques to circumnavigate dephasing of the accelerated electrons in these devices \cite{Palastro2020,Caizergues2020}. 

Currently, generating high-intensity elongated foci requires fixed and often highly customised axioptics.
Integrating such specialized optics into existing experimental setups can be costly, complex and can subsequently restrict the broader functionality of the setup \cite{Liberman2025, Shaw2025}.

\begin{figure}[ht]
\centering
\includegraphics[width=8.6cm]{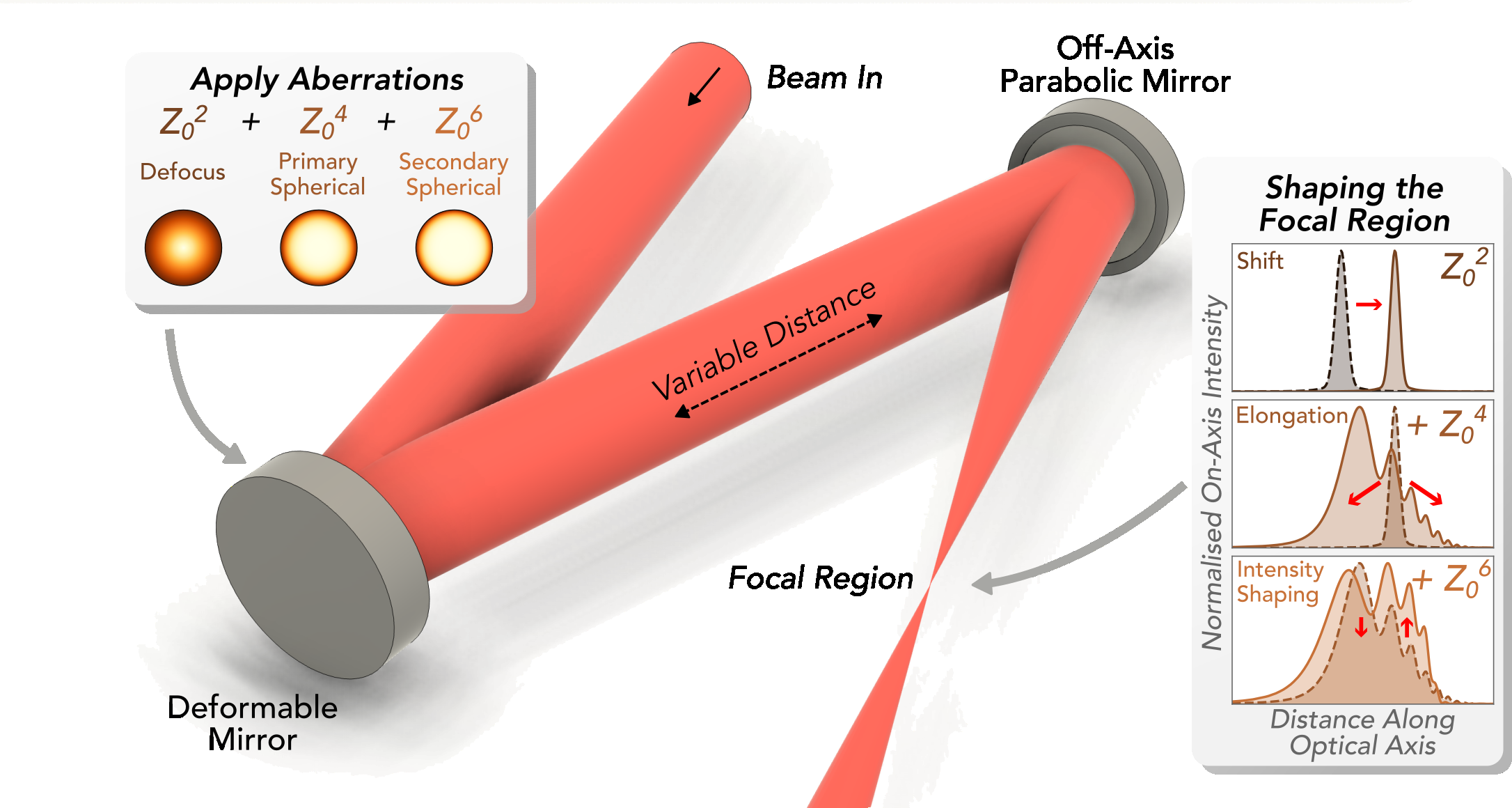}
\caption{\label{fig:conceptfig} Schematic illustrating the separation of tailoring and focusing to achieve an elongated focus. A deformable mirror is used to apply different Zernike aberrations to a laser pulse. After some propagation, an OAP focuses the beam to the focal region. Different combinations of aberrations can be used to independently shape the position, elongation and intensity distribution of the focal region.}
\end{figure}

In this Letter, we propose an accessible and flexible method to generate and programmatically tune such foci. 
The predominant effect of an axioptic is, similar to a standard parabolic mirror, to focus the beam to a region of high-intensity. Provided that the elongation of this region is sufficiently small in comparison to the focal length, the elongation can be treated as a perturbation. 
Indeed, it has previously been noted that an axiparabola predominantly adds spherical aberrations in addition to a focusing phase \cite{Smartsev2019,Palastro2020,Simpson2022,Ambat2023,Pigeon2024}.
Here we explicitly separate focal elongation into two stages, \emph{tailoring} and \emph{focusing}, and propose that these tasks can be performed by two independent optics separated by free-space propagation. Figure \ref{fig:conceptfig} elucidates this concept, showing tailoring of the laser pulse with a deformable mirror and focusing with a standard OAP. 
The two optics are separated by a distance loosely limited by the constraints discussed below.

In the absence of free-space propagation between the optics the total phase, $\Phi$, of a beam after the deformable mirror -- OAP combination can be expressed as
\begin{equation}
\label{eq:totalphase}
    \Phi(r) =  -\frac{k r^2}{2f_0} + \phi(r),
\end{equation}
where the first term is the phase of an OAP with focal length $f_0$, $r$ is the radial coordinate and $k$ is the laser wave vector related to the laser wavelength $\lambda$ as $k=2\pi/\lambda$. 
To preserve the radial symmetry typical of axioptics, we add only radially symmetric aberrations to the pulse, such that
\begin{equation}
\label{eq:zernike}
\phi(r) = \sum_{n=2,4,...} k \alpha_n Z_n^0\left(\frac{r}{R}\right),
\end{equation}
where $Z_n^0$ represent the radially symmetric Zernike polynomials defined over a pupil of radius $R$ and $\alpha_n$ is the root-mean-square (rms) deformation of the laser wavefront in meters. Thus, the total phase can be expanded as a series in $r^{2n}$ with $n\geq0$, encompassing several axioptic phase definitions \cite{Davidson1991, Sochacki1992, Smartsev2019}.
 
Importantly, the two optics can be \emph{spatially separated} along the beam path with the deformable mirror upstream of the focusing optic.
To determine the allowable free-space propagation between the tailoring and focusing stages, while preserving their expected function, we estimate the transverse deformation of the beam due to the aberrations added by the deformable mirror. 
Consider an input beam of characteristic radius $w$ with a flat wavefront. 
In a first step, a deformable mirror imprints a set of radially symmetric Zernike aberrations $\phi(r)$ on the laser wavefront. 
The beam then propagates a distance $z$ along the optical axis where it impinges upon the focusing optic. 
Considered as a geometric optics problem, one can show that during this propagation, a beamlet on the laser wavefront moves transversely to the propagation axis by $|z\, \phi'|/ k$, where $\phi' = d \phi/dr$.
Provided that this movement is small with respect to the beam size, and additionally that diffraction effects of the unaberrated beam are also small over this distance, the spatial beam profile will remain largely unchanged during propagation between the optics.
Mathematically, these conditions can be written as $|z\, \phi'| / k w \ll 1$ and $F > 1 $, where $F = w^2/\lambda z$ is the Fresnel number.
These conditions can be treated as a relatively loose constraint on optic separation to ensure no significant change occurs in the predicted system performance.

A characteristic feature of elongated foci is that different radial positions on the input aperture focus to different longitudinal positions in the focal region. For an optic with a phase given by eq. \ref{eq:totalphase}, one can further use geometric optics to estimate the focal position of the beam as a function of radius on the input aperture \cite{Smartsev2019}:
\begin{equation}
\label{eq:focusposition}
f(r) = \frac{f_0}{1 - \frac{f_0}{kr} \phi^{'}} - \frac{1}{2 k}\left( \phi - \frac{ r}{2} \phi^{'} \right).
\end{equation}
If a defocus aberration $Z_2^0$ is applied, no elongation is observed and the focal position only shifts away from the geometric focus of the parabola by $\Delta f = (4 \sqrt{3} f_0^2 \alpha_2)/(R^2 - 4 \sqrt{3} f_0 \alpha_2).
$
Elongation and intensity tailoring of the focus is achieved by the addition of higher order radially symmetric Zernike aberrations, $Z_4^0$,$Z_6^0$,..., which are shown in an inset in Figure \ref{fig:conceptfig}. 

An important case occurs when the defocus and primary spherical aberrations are added to produce a phase shift proportional to $r^4$ with no $r^2$ dependence.
This is the case when $\alpha_2 = 3 \sqrt{5/3}\, \alpha_4$. The geometric focal extent is then given by

\begin{equation}
\label{eq:length}
L_f = \frac{3 \alpha_{4} \left(\sqrt{5} R^{2} - 120 \alpha_{4} f_{0} + 8 \sqrt{5} f_{0}^{2}\right)}{R^{2} - 24 \sqrt{5} \alpha_{4} f_{0}}.
\end{equation}
In this case, in which the focus is elongated but not shifted, the elongated focus starts at the geometric focus of the OAP and extends downstream for $\alpha_4>0$. 
For eq. \ref{eq:length}, the pupil radius is set equal to the beam radius. 

\begin{figure}[ht]
\centering
\includegraphics[width=3.4in]{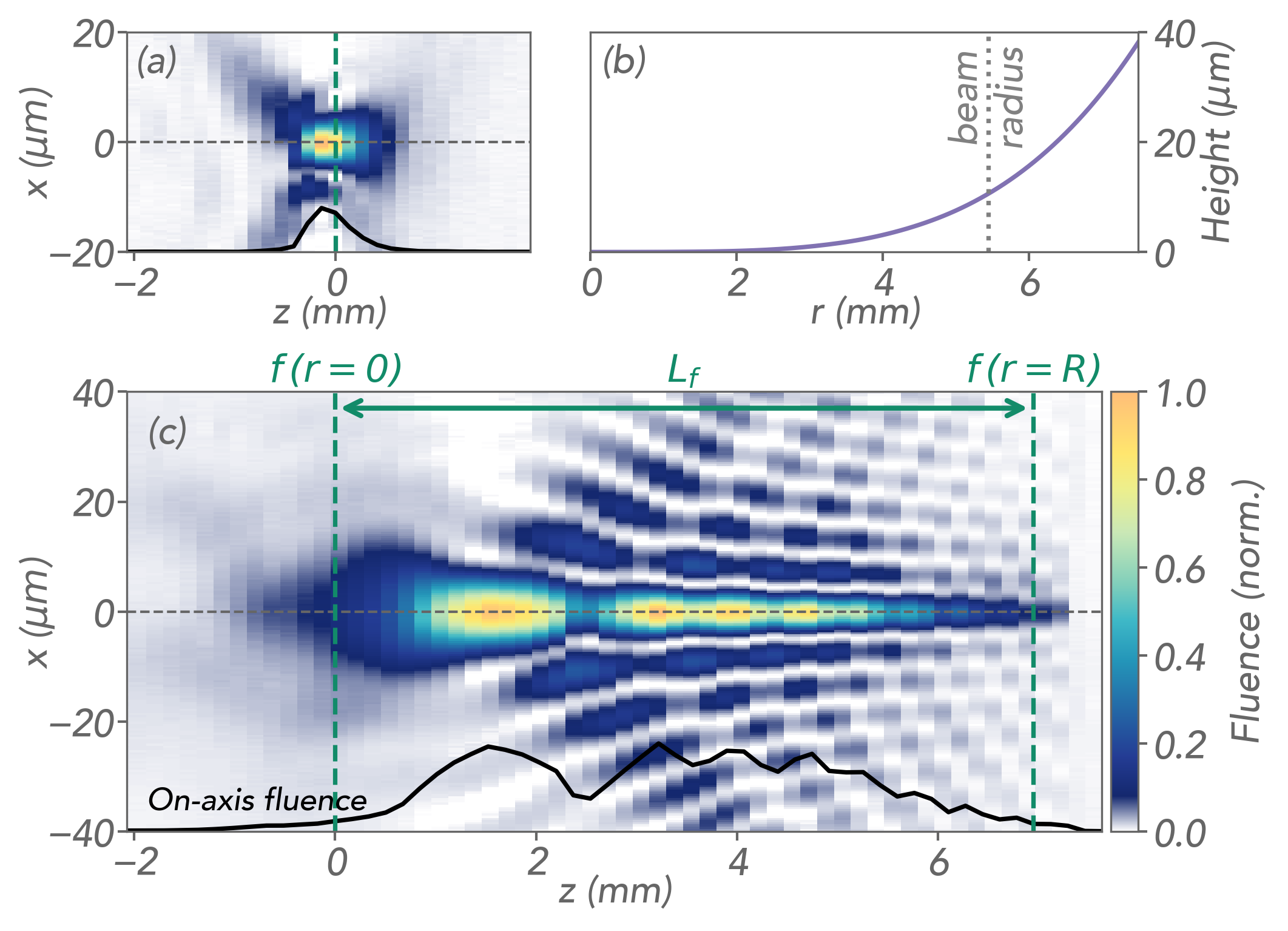}
\caption{\label{fig:phaseplates} (a) Measured focus generated using a $f_0 =$ 100 mm lens. (b) Measured surface profile of the phase plate. (c) Measured elongated focus generated with a phase plate \SI{150}{\mm} prior to the lens. The theoretical predictions of the focal length according to eqs. \ref{eq:focusposition} and \ref{eq:length} are shown in dashed lines.}
\end{figure}

To demonstrate the basic principles above, we performed proof-of-concept experiments using a \SI{780}{\nano\meter} CW laser beam together with a phase plate designed to impart a phase shift proportional to $r^4$. The intensity full-width half-maximum (FWHM) of the beam was \SI{10.9}{\milli\meter} and it was focused using a \SI{100}{\milli\meter} lens to a focal camera setup consisting of a 10x microscope objective and camera mounted on a motorised translation stage. 
The phase plate was a one inch diameter transmissive even-asphere optic made of fused silica. It imparted $\alpha_4 =$ \SI{0.36}{\micro\meter} and $\alpha_2 = 3\sqrt{5/3}(\SI{0.36}{\micro\meter}) = $ \SI{1.39}{\micro\meter} over the beam diameter.
The phase plate was placed approximately \SI{150}{\milli\meter} from the lens, to limit beam evolution during the propagation (max. $|z\, \phi'| / k w  \sim 0.1$).  
Fig. \ref{fig:phaseplates} (a) and (b) show respectively the unaberrated focus and the surface profile of the central region of the phase plate. Upon addition of the phase plate, the focus elongates, with fig. \ref{fig:phaseplates} (c) showing the resulting elongated focal region together with the analytic prediction of the focal extent, eqs. \ref{eq:focusposition} and \ref{eq:length}. The experiment and theory show excellent agreement. 

Beyond simply elongating the focal region, many applications also require control of the longitudinal intensity profile. 
For the concept presented here, the longitudinal intensity distribution of the elongated focus depends upon the spatial intensity profile of the beam at the surface of the tailoring optic, the aberrations added to the beam, the amount of free-space propagation between the tailoring optic and the focusing optic, and the focusing optic itself.
For a fixed input spatial intensity profile, fixed focusing optic, and a fixed separation between the optics, a tailored phase comprising a combination of primary, secondary etc. spherical aberrations may be used to redistribute the energy within the focal region and thus shape the longitudinal intensity profile. This phase tailoring task is well suited to programmable optics such as deformable mirrors. 

In the following section we explore this concept further, in particular the use of deformable mirror to programmatically add radially symmetric Zernike aberrations to elongate and shape the longitudinal intensity profile of the focus of an existing high-power laser system used for laser plasma acceleration. These experiments were conducted at the LUX facility \cite{Delbos2018,Maier2020}. 

At LUX, an \emph{Imagine Optic ILAO-Star 200} deformable mirror with 52 actuators is located just upstream of the final compressor of the Ti:Sapphire laser system, ANGUS, which can deliver up to \qty{2.6}{\joule} of energy in a \qty{39}{\femto\second} FWHM pulse centered on \qty{800}{\nano\meter} in a beam with an order 3.4 flattened Gaussian profile with a \qty{68}{\milli\meter} FWHM \cite{Santarsiero1997}. After compression the laser travels along a non-imaging beam transport to the interaction chamber where it is focused by a \qty{2}{\meter} OAP. The total distance between the deformable mirror and the OAP is approximately \qty{34}{\meter}.

\begin{figure}[t]
\includegraphics[width=8.6cm]{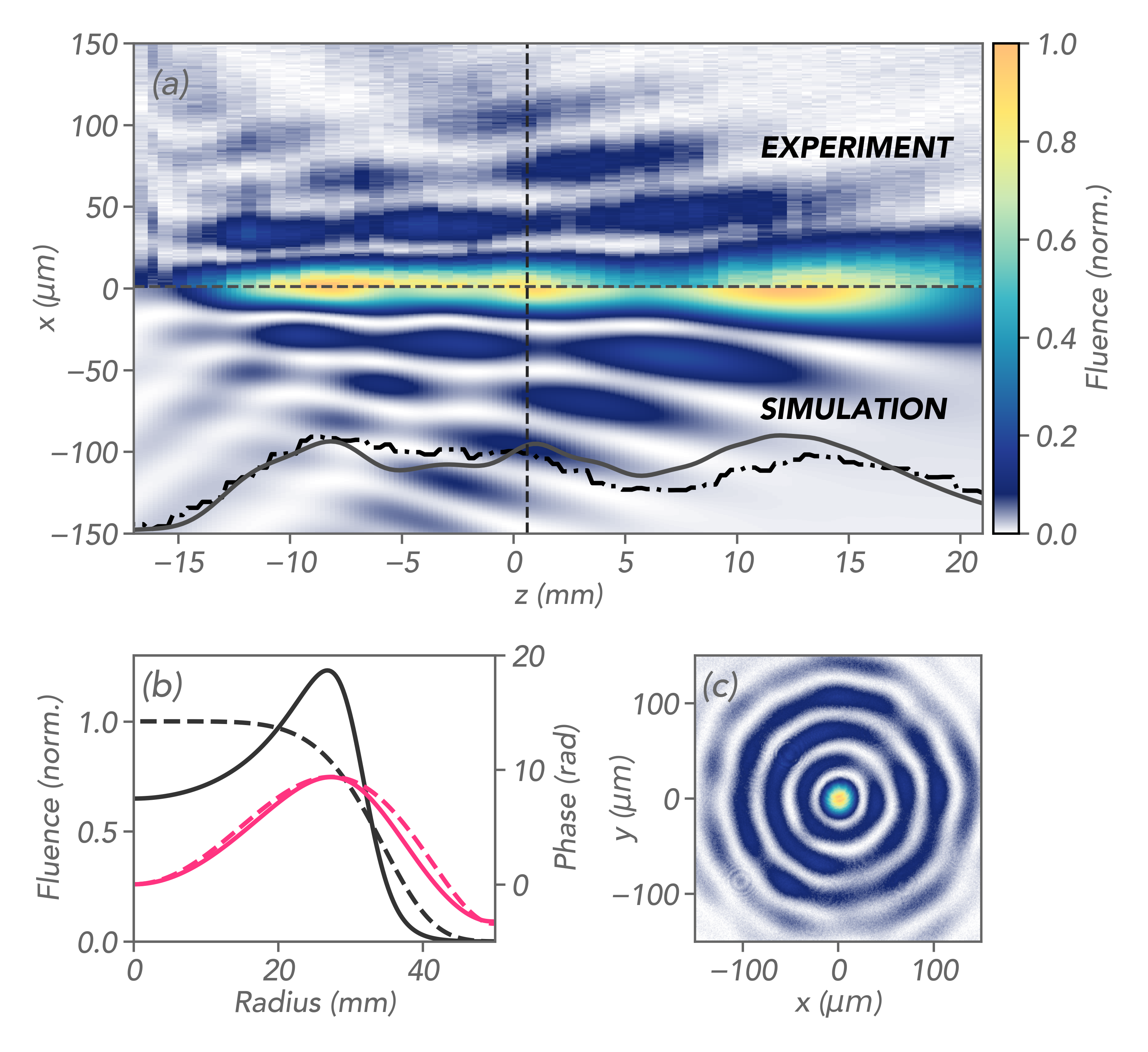}
\caption{\label{fig:luxFocalscans} (a) Fluence profile in the focal region of the LUX accelerator. The upper half of the image is the experimentally measured data from a focal scan of the aberrated beam while the lower half is a simulation, performed with LASY, in which the experimentally set values for the deformable mirror have been added to the pulse. In the bottom half, the on-axis fluence for both experiment (dashed dotted) and simulation (solid) are shown. (b) The simulated near field fluence (black) and phase (magenta) profiles at the deformable mirror (dashed) and at the OAP (solid). (c) a transverse slice of the measured focal intensity profile taken along the vertical dashed line in (a).
}
\end{figure}

The wavefront of the attenuated pulsed beam is flattened prior to compression by the deformable mirror to facilitate the generation of a high-quality focus, with a measured FWHM intensity spot size of \SI{22}{\micro\meter} and Rayleigh length of \SI{2.2}{\milli\meter}. 
Starting from this flattened wavefront: defocus, primary spherical and secondary spherical aberrations were added to the pulse resulting in the generation of an elongated focus, more than \SI{35}{\milli\meter} long, as can be seen in figure \ref{fig:luxFocalscans}. LASY \cite{lasy} was used to perform simulations of the optical system from the deformable mirror to focus, including the \qty{34}{m} of beam propagation. Angular spectrum method and resampling Fresnel propagators were employed for free-space propagation and focusing respectively. Good agreement was found between the measured data and simulations.

In comparison to the phase plate measurements above, here there is additionally secondary spherical aberration. 
In this case the defocus was used to approximately centre the elongated focal region around $z=0$ which corresponds to the geometric focus of the unaberrated system.
The secondary spherical aberration is used to redistribute energy within the focal region, flattening the longitudinal intensity profile. 
As mentioned above, the free-space propagation can also influence the longitudinal intensity profile. 
Here, it is observed that while laser energy is redistributed transversely --- leading to a modified spatial intensity profile of the beam at the OAP surface compared to the surface of the deformable mirror --- the phase profile does not significantly evolve. For the conditions shown, the maximum of $|z\, \phi'| / k w$ is also  $\sim 0.1$.

The results here demonstrate the ability to decompose the production of an elongated focus into two stages separated by free-space propagation, allowing independent optical elements to perform each step. 
While in the first example a custom phase plate was used for phase tailoring, it is clear that, due to nonlinear phase acrual, such optics do not scale well to use at high laser intensities. 
In such cases, the deformable mirror offers a clear advantage due to its ability to operate at high pulse intensities and to programmatically modify the imparted phase and thus the longitudinal intensity profile.

It is worth noting here that in the second example, the magnitude of aberrations added were within the dynamic range of a standard deformable mirror setup, even after flattening of the wavefront. Specifically, a defocus of $\alpha_2 =$ \SI{-0.2}{\micro\meter}, a primary spherical of $\alpha_4 =$ \SI{-0.35}{\micro\meter}, and a secondary spherical of $\alpha_6 =$ \SI{0.1}{\micro\meter} were added to the pulse, with an aberration pupil radius of \SI{45}{\milli\meter}. 
Similar deformable mirror systems are commonplace in many high-power laser facilities worldwide, meaning that the techniques described here could be readily applied to a variety of facilities globally. 

In general, the extension of the focal region can be limited by three factors; the phase which can be added by the deformable mirror, the focusing properties of the unaberrated beam (focal length and beam size), and the distance between the deformable mirror and OAP. In the case of typical LPA-like setups, the limiting factor tends to be the constraints arising from the propagation distance between the deformable mirror and the OAP. 
With reasonable limits on these constraints as described above, one can comfortably produce regions of high-intensity up to several percent of the focal length of the focusing optic. 
In the following we explore how this concept could be leveraged to support applications in multi-GeV electron acceleration and dephasingless wakefield acceleration.

The formation of a Hydrodynamic Optical-Field-Ionized (HOFI) plasma waveguide \cite{Shalloo2018}, used in multi-GeV LPAs \cite{Oubrerie2022-2,Miao2022b,Picksley2023,Picksley2024}, typically requires an auxiliary beamline with a separate focusing geometry. 
However, relative pointing fluctuations between the waveguide-forming and drive-pulses remains a significant source of instability \cite{Picksley2023}.
The dual-optic technique described here could allow the HOFI pulse and main drive pulse to utilise a common final focusing optic, potentially alleviating relative pointing jitter due to mechanical vibrations and enabling the generation of a stable guiding structure. 
One facility which would benefit from such developments is the proposed \SI{6}{\giga\electronvolt} Plasma Injector for PETRA IV \cite{pip4cdr}. 
The generation of a suitably long region of high laser intensity is well within the practical limitations of our technique, although one would also need to consider the impact of ionization-induced refractive effects which can additionally modify the length of the focal region.  

\begin{figure}
\includegraphics[width=8.6cm]{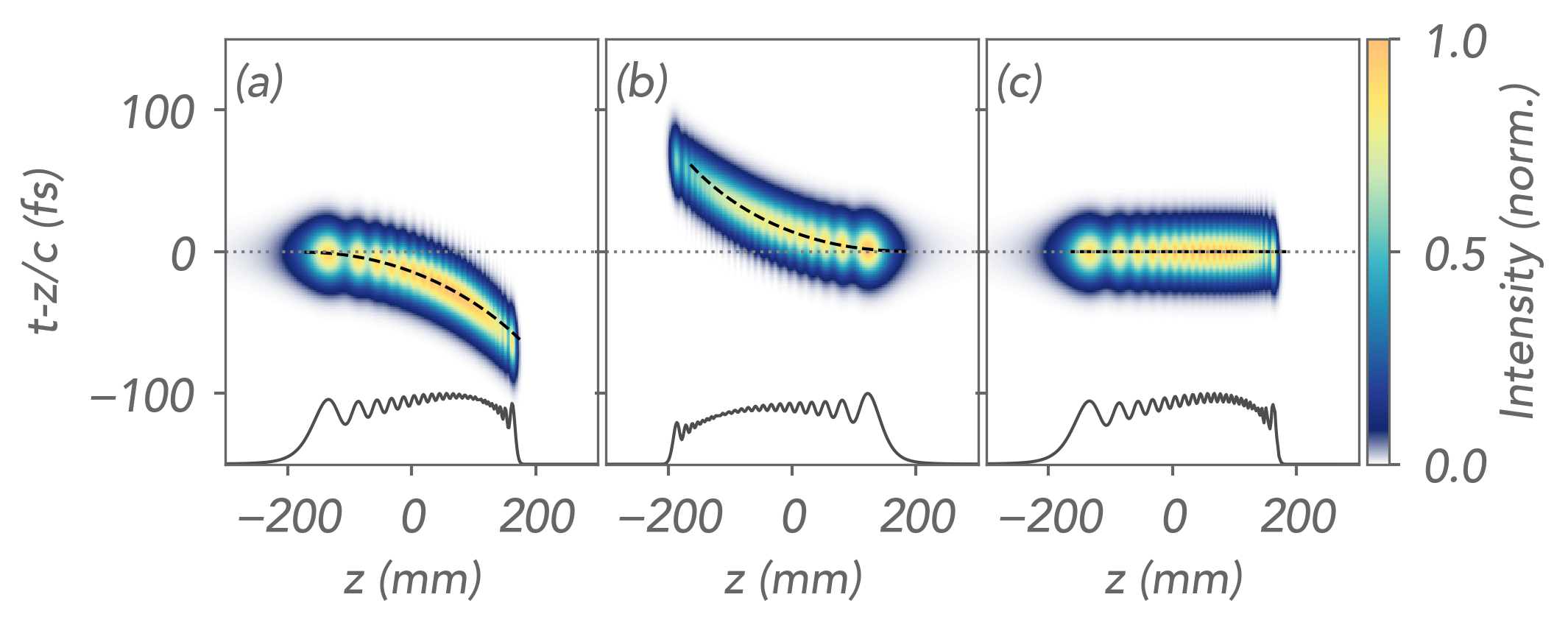}
\caption{\label{fig:flyingfocus} Spatio-temporal structure of elongated foci. In all cases, an \SI{800}{\nano\meter}, \SI{30}{\femto\second} pulse with a super Gaussian transverse profile of \SI{200}{\milli\meter} diameter is tailored and then propagated \SI{20}{\meter} before being focused by a $f_0 = $ \SI{6}{\meter} OAP. In (a) a phase $\phi$ is applied with $(\alpha_2,\alpha_4,\alpha_6)=(1.8, 1.8, .22)$ \si{\micro\meter} and $|z\, \phi'| / k w = 0.1$. In (b) $-\phi$ is applied. In both cases, superluminal propagation of the intensity peak is observed. The dashed line shows the analytic prediction of the arrival time, which shows excellent agreement, despite tailoring being performed prior to propagation. The solid line at the bottom of each plot shows the axial intensity lineout. In (c) the same phase $\phi$ and a radial group delay $c\tau(r)$ are added to achieve luminal propagation. 
}
\end{figure}

In the context of dephasingless laser wakefield acceleration \cite{Caizergues2020,Palastro2020}, it is advantageous to tune the velocity of the intensity peak along the optical axis. 
Indeed, it can be shown that in addition to elongating the focus along the optical axis, adding the phase described in eq. \ref{eq:zernike} impacts the spatio-temporal behaviour of the pulse, leading to an axial intensity peak which travels superluminally \cite{Ambat2023}, a flying focus \cite{Sainte-Marie2017, Froula2018}. 
The time of arrival $t(r)$ of the rays at the focus can be expressed as

\begin{equation}
\label{eq:arrivaltime}
    c t(r) = \frac{f_0}{1 - \frac{f_0}{kr} \phi^{'}} + \frac{1}{2 k}\left( \phi - \frac{ r}{2} \phi^{'} \right).
\end{equation}
The velocity of the intensity peak on axis can be calculated from eqs. \ref{eq:focusposition} and  \ref{eq:arrivaltime} as $v_\textrm{ff}/c = (\partial{f}/\partial{r}) / (\partial{ct}/\partial{r})$. 
The time of arrival can be tuned by addition of a radial group delay $c \tau(r)$ as shown in fig. \ref{fig:flyingfocus}. Here, a radial group delay is added to generate an elongated focus which travels luminally, $v_\textrm{ff}/c = 1$.
While the radial group delay may be tailored by a fixed echelon optic, a programmatic solution based on an adaptive optic system was recently proposed \cite{Ambat2023}. 
One could then consider an all-adaptive-optic solution to programmable control of high-intensity flying foci. 
Similar to the elongation of the focus, the temporal shaping can occur far from the final focusing optic (as demonstrated in fig. \ref{fig:flyingfocus}) and could thus be easily integrated in an existing setup.

In summary, we have presented a novel technique for the programmatic elongation and intensity shaping of high-power laser pulses using a deformable mirror in combination with a standard focusing optic. 
The use of two independent optics for tailoring and focusing provides both flexibility of implementation and tunability of the focal intensity profile. 
The ability to spatially separate these processes, by relatively large distances, enables the technique to be easily applied to a variety of existing systems. 
We presented results with a custom phase plate demonstrating, in a simple optical setup, the ability to separate the tailoring and focusing processes.
The length of the measured elongated focus agrees well with theoretical calculations derived using geometric optics.
Further, we have shown that the amount of spherical aberration required to significantly elongate the focus can be readily provided by existing deformable mirror systems, enabling programmatic control over the focal properties. 
This was demonstrated at an existing LPA facility, LUX.
Here, the focus was elongated to well beyond the Rayleigh length with the existing deformable mirror installed before the compressor. The intensity profile was further tailored by carefully tuning the relative amounts of primary and secondary spherical aberrations. 
Our experimental results are supported by a theoretical framework and by optical simulations with LASY \cite{lasy}.
Finally, we considered how these techniques might extend to two important applications, namely multi-GeV laser-driven plasma acceleration and the generation of flying foci for dephasingless wakefield acceleration.

\begin{backmatter}
\bmsection{Funding}
This work was supported by funding from the Deutsche Forschungsgemeinschaft (DFG, German Research Foundation) – Project number 531352484.

\end{backmatter}

\providecommand{\noopsort}[1]{}\providecommand{\singleletter}[1]{#1}%


\begin{thebibliography}{10}
\newcommand{\enquote}[1]{``#1''}

\bibitem{McLeod1954}
J.~H. McLeod, {\protect\JournalTitle{J. Opt. Soc. Am.}} \textbf{44}, 592 (1954).

\bibitem{Davidson1991}
N.~Davidson, A.~A. Friesem, and E.~Hasman, {\protect\JournalTitle{Opt. Lett.}} \textbf{16}, 523 (1991).

\bibitem{Sochacki1992}
J.~Sochacki, S.~Bar\'{a}, Z.~Jaroszewicz, and A.~Ko{\l}odziejczyk, {\protect\JournalTitle{Opt. Lett.}} \textbf{17}, 7 (1992).

\bibitem{Smartsev2019}
S.~Smartsev, C.~Caizergues, K.~Oubrerie, \emph{et~al.}, {\protect\JournalTitle{Opt. Lett.}} \textbf{44}, 3414 (2019).

\bibitem{Oubrerie2022}
K.~Oubrerie, I.~A. Andriyash, R.~Lahaye, \emph{et~al.}, {\protect\JournalTitle{Journal of Optics}} \textbf{24}, 045503 (2022).

\bibitem{Miao2022}
B.~Miao, L.~Feder, J.~E. Shrock, and H.~M. Milchberg, {\protect\JournalTitle{Opt. Express}} \textbf{30}, 11360 (2022).

\bibitem{Picksley2024}
A.~Picksley, J.~Stackhouse, C.~Benedetti, \emph{et~al.}, {\protect\JournalTitle{Phys. Rev. Lett.}} \textbf{133}, 255001 (2024).

\bibitem{Palastro2020}
J.~P. Palastro, J.~L. Shaw, P.~Franke, \emph{et~al.}, {\protect\JournalTitle{Phys. Rev. Lett.}} \textbf{124}, 134802 (2020).

\bibitem{Caizergues2020}
C.~Caizergues, S.~Smartsev, V.~Malka, and C.~Thaury, {\protect\JournalTitle{Nature Photonics}} \textbf{14}, 475 (2020).

\bibitem{Liberman2025}
A.~Liberman, A.~Golovanov, S.~Smartsev, \emph{et~al.}, {\protect\JournalTitle{arXiv}}  (2025).

\bibitem{Shaw2025}
J.~L. Shaw, M.~V. Ambat, K.~G. Miller, \emph{et~al.}, {\protect\JournalTitle{Physics of Plasmas}} \textbf{32}, 083107 (2025).

\bibitem{Simpson2022}
T.~T. Simpson, D.~Ramsey, P.~Franke, \emph{et~al.}, {\protect\JournalTitle{Opt. Express}} \textbf{30}, 9878 (2022).

\bibitem{Ambat2023}
M.~V. Ambat, J.~L. Shaw, J.~J. Pigeon, \emph{et~al.}, {\protect\JournalTitle{Opt. Express}} \textbf{31}, 31354 (2023).

\bibitem{Pigeon2024}
J.~J. Pigeon, P.~Franke, M.~L.~P. Chong, \emph{et~al.}, {\protect\JournalTitle{Opt. Express}} \textbf{32}, 576 (2024).

\bibitem{Delbos2018}
N.~Delbos, C.~Werle, I.~Dornmair, \emph{et~al.}, {\protect\JournalTitle{Nuclear Instruments and Methods in Physics Research Section A: Accelerators, Spectrometers, Detectors and Associated Equipment}} \textbf{909}, 318 (2018). 3rd European Advanced Accelerator Concepts workshop (EAAC2017).

\bibitem{Maier2020}
A.~R. Maier, N.~M. Delbos, T.~Eichner, \emph{et~al.}, {\protect\JournalTitle{Phys. Rev. X}} \textbf{10}, 031039 (2020).

\bibitem{Santarsiero1997}
R.~B. M.~Santarsiero, D.~Aiello and S.~Vicalvi, {\protect\JournalTitle{Journal of Modern Optics}} \textbf{44}, 633 (1997).

\bibitem{lasy}
M.~Thévenet, I.~A. Andriyash, L.~Fedeli, \emph{et~al.}, \enquote{Lasy: Laser manipulations made easy,}  (2024).

\bibitem{Shalloo2018}
R.~J. Shalloo, C.~Arran, L.~Corner, \emph{et~al.}, {\protect\JournalTitle{Phys. Rev. E}} \textbf{97}, 053203 (2018).

\bibitem{Oubrerie2022-2}
K.~Oubrerie, A.~Leblanc, O.~Kononenko, \emph{et~al.}, {\protect\JournalTitle{Light: Science \& Applications}} \textbf{11}, 180 (2022).

\bibitem{Miao2022b}
B.~Miao, J.~E. Shrock, L.~Feder, \emph{et~al.}, {\protect\JournalTitle{Phys. Rev. X}} \textbf{12}, 031038 (2022).

\bibitem{Picksley2023}
A.~Picksley, J.~Chappell, E.~Archer, \emph{et~al.}, {\protect\JournalTitle{Phys. Rev. Lett.}} \textbf{131}, 245001 (2023).

\bibitem{pip4cdr}
I.~Agapov, S.~Antipov, R.~Brinkmann, \emph{et~al.}, \emph{{T}he {P}lasma {I}njector for {PETRA} {IV}: {E}nabling {P}lasma {A}ccelerators for {N}ext-generation {L}ight {S}ources. {C}onceptual {D}esign {R}eport} (Deutsches Elektronen-Synchrotron DESY, Hamburg, 2025).

\bibitem{Sainte-Marie2017}
A.~Sainte-Marie, O.~Gobert, and F.~Qu\'{e}r\'{e}, {\protect\JournalTitle{Optica}} \textbf{4}, 1298 (2017).

\bibitem{Froula2018}
D.~H. Froula, D.~Turnbull, A.~S. Davies, \emph{et~al.}, {\protect\JournalTitle{Nature Photonics}} \textbf{12}, 262 (2018).

\end{thebibliography}
\end{document}